\documentclass[12pt,a4paper,tightenlines,nofootinbib]{revtex4}

\usepackage{amsmath}
\usepackage{amsfonts}
\usepackage{amssymb}
\usepackage{epsf}

\begin{document}

\author{Ariel M\'egevand$^{1}$ and Francisco Astorga$^{2}$}
\affiliation{$^1$ IFAE, Universitat Aut{\`o}noma de Barcelona,
08193 Bellaterra (Barcelona), Spain} \affiliation{$^2$ Instituto
de F\'{\i}sica y Matem\'aticas, Universidad Michoacana de San
Nicol\'as de Hidalgo, Edificio C--3 Cd. Universitaria, A. Postal
2--82, 58040 Morelia, Michoac\'an, Mexico}
\title{Generation of baryon inhomogeneities in the
electroweak phase transition}
\date{\today}

\begin{abstract}

It is known that baryon number inhomogeneities may arise as a
consequence of electroweak baryogenesis. Their geometry, size, and
amplitude depend on the parameters that characterize the
baryogenesis mechanism, as well as on those that determine the
phase transition dynamics. We investigate  this parametric
dependance. We show that in the case of the minimal supersymmetric
standard model, the geometry of the inhomogeneities most probably
consists of spherical regions of high density surrounded by
low-density walls, in contrast to the case of the minimal standard
model. In this supersymmetric extension we find that density
contrasts of up to a factor of 100 may arise. This amplitude
increases for higher values of the latent heat or lower values of
the bubble wall tension, and can be significantly larger in
different extensions of the standard model. Such inhomogeneities
may thus affect the dynamics of the subsequent quark-hadron phase
transition.
\end{abstract}

\maketitle

\section{Introduction}

The generation of inhomogeneities in the baryon number density is
a possible outcome of the QCD and the electroweak phase
transitions \cite{w84,inhomQCD,h95,bdr95}. These may arise as a
consequence of the perturbations produced by the walls of
expanding bubbles in the surrounding plasma. In the QCD case,
baryons, which are much heavier in the hadron phase than in the
deconfined quark phase, are pushed away by the walls of expanding
bubbles as the phase transition develops. Then, as the volume
occupied by the hadron phase grows at the expense of the quark
phase, baryons are driven into small regions of space \cite{w84}.
Hence, it is believed that the general geometry of the QCD
inhomogeneities is that of localized clumps of high density
surrounded by voids of low density.

In the electroweak phase transition, baryon number inhomogeneities
can arise as a consequence of the generation of the baryon
asymmetry of the Universe (BAU), whether  the latter is produced
through the standard mechanism of electroweak baryogenesis
\cite{h95} or in electroweak cosmic strings \cite{bdr95}. In
general, the electroweak phase transition provides the three
Sakharov conditions \cite{s67} for baryogenesis, namely, baryon
number violation, $C$ and $CP$ violation, and a departure from
thermal equilibrium. These conditions must combine in such a way
as to produce a net difference between number densities of baryons
and antibaryons. Interestingly, the idea of electroweak
baryogenesis requires physics beyond the minimal standard model
(SM) in order to give a quantitatively successful result. The
basic idea is the following (for reviews on electroweak
baryogenesis see \cite{ckn93}).

The standard mechanism assumes a first-order phase transition in order to
achieve the non-equilibrium condition. As bubbles expand, a chiral flux is
injected in front of their walls due to $CP$ violating interactions of the
bubble walls with the particles of the plasma. Thus, an asymmetry between
left handed quarks and their antiparticles is generated near the interfaces.
This asymmetry biases the baryon number violating sphaleron processes in the
symmetric phase. The resulting baryon asymmetry is caught by the walls and
enter the bubbles. In order to avoid the washout of the generated BAU when
equilibrium is restored, the sphaleron processes must be suppressed in the
broken symmetry phase. This requirement imposes a condition on the value of
the Higgs field $\phi $ inside the bubbles \cite{s87},
\begin{equation}
\phi _{m}\left( T\right) /T\gtrsim 1\ .  \label{washout}
\end{equation}
Here, $\phi _{m}$ is the vacuum expectation value (VEV) of $\phi $
at temperature $T$, which corresponds to the global minimum of the
free energy. It plays the role of an order parameter, and the
condition (\ref{washout}) states that the phase transition must be
strongly first-order. This usually constrains the parameters of
the theory.

The generated BAU has a strong dependence on the velocity $v_{w}$
of bubble walls. On one hand, if the velocity is too large, the
left-handed density perturbation will pass so quickly through a
given point in space that sphaleron processes will not have enough
time to produce baryons. Thus, the resulting baryon number density
$n_{B}$ will be small in this case. On the other hand, for very
small velocities thermal equilibrium will be restored and the
baryon asymmetry will be erased by sphalerons. Hence, the baryon
production will be small again. Therefore, the generated baryon
number has a maximum for a certain wall velocity
$v_{w}=v_{\mathrm{peak}}$, which can be estimated by comparing the
baryon number violation time scale with the time of passage of the
chiral asymmetry \cite{lmt92,ckn92,h95}. Such estimates give a
small value $v_{\mathrm{peak}}\sim 10^{-2}$, which is confirmed by
numerical calculations \cite{ck00,cmqsw01}.

This dependance of $n_{B}$ on the bubble wall velocity is
important for electroweak baryogenesis, since it establishes a
preference for models in which the velocity is close to
$v_{\mathrm{peak}}$. On the other hand, the velocity of bubble
expansion is not constant throughout the phase transition
\cite{h95,dynamics,ariel03,ariel01}. The variation of $ v_w$
causes a variation of the local baryon density left behind by the
walls along the bubble radius. It is thus evident that generating
the observed BAU in the electroweak phase transition entails the
formation of inhomogeneities in the baryon number density. These
inhomogeneities are spherically symmetric and centered at the
nucleation points.

A general feature of first-order phase transitions is the
slow-down of bubble expansion due to the release of latent heat
\cite{ariel03}. In the case of the electroweak phase transition,
it is known that the velocity of bubble walls may decrease from an
initial value $ v_{i}\sim 10^{-1}-10^{-2}$ to a minimum velocity
$v_{m}\sim 10^{-3}-10^{-4}$, but the variation can be larger,
depending on the model \cite{h95,ariel01}.  The maximum variation
of the baryon density $ n_{B}\left( v_{w}\right) $ as $v_{w}$
decreases from $v_{i}$ to $v_{m}$ determines the amplitude of the
baryon inhomogeneities generated inside each bubble. The exact
profile $n_B(r)$ along the radial direction and the general
geometry of the inhomogeneities depend on the evolution of
$v_{w}(t)$. For instance, if the initial velocity $v_{i}$ is close
to the peak velocity $v_{\mathrm{peak}}$, then clumps of high
baryon density will be formed, surrounded by voids of low density.
On the contrary, if $v_{m}\simeq v_{\mathrm{peak}}$, walls of high
density will surround voids of low density. The size of the
inhomogeneities is roughly given by the mean separation between
the centers of bubbles at the end of the phase transition.

The survival of the inhomogeneities at later epochs depends on the
amplitude of the fluctuations as well as on the separation of the
centers of fluctuations, which must be larger than the diffusion
length of baryons for that fluctuation amplitude. In general, the
inhomogeneities will have different cosmological consequences,
depending on the epoch in which the phase transition occurs.
Baryon inhomogeneities generated in the QCD phase transition have
been intensively investigated due to their influence on Big Bang
Nucleosynthesis (BBN) \cite{w84,inhomQCD,ahs87}. On the contrary,
baryon inhomogeneities generated at the electroweak scale are not
expected to survive long enough to affect BBN \cite{h95,bdr95}.

However, electroweak inhomogeneities may survive until the QCD
scale \cite{s03,jf94}. Then, the dynamics of the quark-hadron
phase transition may be affected by their presence
\cite{s03,cm96,h83}. This is because the critical temperature
$T_c$ of the quark-hadron phase transition depends on the chemical
potential, and therefore on the number density, being $T_{c}$
lower in regions with higher baryon density \cite{hjssv98,Fodor}.
Hence, the process of bubble nucleation is delayed in such
regions. The effect of pre-existing baryon inhomogeneities on the
quark-hadron phase transition depends on the profile of the
inhomogeneities.

Assume for instance that the geometry is that of small regions
with low baryon density in a background of higher density. Then,
since the critical temperature is reached first in the low-density
regions, the inhomogeneities may act as impurity sites where
hadron-phase bubbles can nucleate before the supercooling required
for spontaneous nucleation is reached \cite{ariel03}. Thus, the
QCD phase transition may proceed by inhomogeneous nucleation at
$T_c$, rather than by supercooling and homogeneous nucleation with
a rate $\Gamma$ per unit time and volume. Therefore, the size
scale of the inhomogeneities determines the mean separation
between centers of nucleation. This distance, in turn, sets the
scale of the QCD inhomogeneities.

If, on the contrary, the inhomogeneities have the form of small
lumps of high baryon density surrounded by a background of lower
density, then the phase transition will be delayed inside the
lumps. Since the Universe quickly reheats close to $T_{c}$, the
high density regions stay in the quark-gluon plasma phase, while
the hadron-phase bubbles grow outside. This also affects the QCD
generation of inhomogeneities, since baryon number tends to
accumulate in the deconfined quark phase, as explained above. As a
consequence, inhomogeneities generated at the electroweak scale
may be amplified at the QCD phase transition, and therefore
indirectly affect primordial nucleosynthesis \cite{s03}.

In this paper we study the formation of baryon density
inhomogeneities in a baryon-generating electroweak phase
transition. This issue has been previously investigated by Heckler
\cite{h95} for the minimal standard model. However, as we already
mentioned, electroweak baryogenesis is possible only in extensions
of the SM. The function $n_B\left(v_w\right)$, as well as the
dynamics of the phase transition, can be quite different in such
extensions. As a consequence, the kind of inhomogeneities that
form may be quantitatively as well as qualitatively different. In
particular, in Ref. \cite{h95} the wall velocity was supposed to
lie well on the right of $v_{\mathrm{peak}}$, and a dependance
$n_{B}\propto v_{w}^{-1}$ was assumed. In contrast, recent
calculations \cite{ck00,cmqsw01,js01} show that the initial
velocity is likely to be around $v_{\mathrm{peak}}$, at least in
the case of the minimal supersymmetric standard model (MSSM). In
such a case, the velocity slow-down may occur completely on the
left of the peak \cite{ariel01} and, as we shall see, yield the
opposite behavior, $n_{B}\propto v_{w}$.

In sec. \ref{baryo} we analyze the dependance of $n_{B}$ on
$v_{w}$ in the whole range of (non-relativistic) wall velocities.
In sec. \ref{dynamics} we study the evolution of the wall velocity
using analytic approximations which allow to determine the
parametric dependance of the amplitude and shape of the
inhomogeneities. Unfortunately, such approximations give only a
qualitative picture, so we also make a more numerical
investigation of the electroweak phase transition: In section
\ref{numerical} we derive appropriate equations for the evolution
of temperature and fraction of volume occupied by the
broken-symmetry phase. We integrate this set of equations
numerically and calculate the baryon number density. In section
\ref{results} we present the results of this calculation. We
consider different values of the parameters, which include the
case of the MSSM. Finally, in section \ref{conseq} we discuss the
possibility that the electroweak baryon inhomogeneities have an
effect on the QCD phase transition. We argue that this is possible
in some extensions of the SM. Our conclusions are summarized in
section \ref{conclu}.

\section{Dependance of electroweak
baryogenesis on the wall velocity \label{baryo}}

The baryon asymmetry originates from the chiral quark asymmetry in
the symmetric phase that is generated as a consequence of the CP
violating currents coming from the wall. The density of
left-handed particles in front of the bubble walls can be
calculated  by considering a set of coupled diffusion equations of
the form (see e.g. \cite{ckn94,hn96})
\begin{equation}
\dot{n}_{i}=D_{i}\nabla ^{2}n_{i}-\sum \Delta _{ai}\Gamma _{a}
\frac{\Delta _{aj}n_{j}}{k_{j}}+\gamma _{i} , \label{difusgral}
\end{equation}
for all particle species $i$ that have $CP$ violating interactions
with the bubble wall. Here, $D_{i}$ are the diffusion constants,
$\Gamma _{a}$ are the rates of reactions that change the number of
particle $i$ by an amount $\Delta _{ai}$ (the corresponding rates
per unit volume are $T^{3}\Gamma _{a}/6$), $ k_{i}$ are
statistical factors: $k_{i}=g_{i}$ for fermions, $k_{i}=2g_{i}$
for bosons, where $g_{i}$ is the number of degrees of freedom of
species $i$, and $\gamma _{i}$ are the rates of generation of
axial numbers at the wall, which depend on the Higgs profile.

In general, the diffusion equations (\ref{difusgral}) must be
solved numerically. However, several approximations can be made in
order to simplify the problem \cite{ck00,cmqsw01,hn96,mr00}. For
instance, the diffusion constants for all quarks can be assumed to
be nearly the same, and some of the rates $\Gamma _{a}$ (e.g.,
those corresponding to strong sphalerons and trilinear Higgs-quark
interactions) can be assumed to be fast enough to be in thermal
equilibrium. As a consequence, several of the particle densities
can be eliminated algebraically, leaving only a few equations to
deal with. In order to avoid entering into the details of any
specific model, we will assume that all the $ n_{i}$ can be
written in terms of just one of them, so only one diffusion
equation needs to be solved (see e.g. \cite{hn96}). This will
allow us to find an analytical solution for the baryon density,
which can be easily inserted in the numerical calculation of the
evolution of the phase transition.

Therefore, we can write a diffusion equation for the density of
left-handed fermions with effective values of $D$, $\Gamma $, and
$\gamma $,
\begin{equation}
Dn_{L}^{\prime \prime }+v_{w}n_{L}^{\prime }-\Gamma n_{L}+\gamma
=0, \label{difus}
\end{equation}
where we have made the usual assumption of a planar wall which
moves non-relativistically to the right, so $\gamma _{i}$ and $
n_{i}$ depend only on $z-v_{w}t$. The diffusion constant $D$ for
the chiral asymmetry depends on the particle spectrum of the
specific model. In supersymmetric extensions of the SM, the main
contributions to $D$ come from the Higgs sector and give $D\sim
100T^{-1}$ \cite{hn96}. The effective axial quark number
relaxation rate $\Gamma $ receives contributions from the helicity
flip due to the top quark mass and from Higgs self-interactions.
In the wall frame, it can be approximated by
\cite{ck00,cmqsw01,hn96} $\Gamma \left( z\right)
=\tilde{\Gamma}\theta \left( -z\right) $, where $\theta $ is
Heaviside's function and $\tilde{\Gamma}\sim 10^{-1}T$. The weak
sphaleron processes are very slow, so they can be disregarded for
relatively large wall velocities. However, we will consider a
range of velocities that may include very small values of $v_{w}$.
Therefore, we should take into account that the effective rate
$\Gamma $ in the symmetric phase is not zero but of the order of
the weak sphaleron rate. Consequently, we will assume that
$\Gamma=\tilde{\Gamma}$ in the broken-symmetry phase, and $\Gamma
=a\Gamma _{ws}$ in the symmetric phase, where $a$ is a numerical
constant that depends on the degrees of freedom contributing to
Eq.~(\ref{difus}), and $\Gamma _{ws}
\sim 10^{-6}T$ \cite{mr00}.

Thus, Eq. (\ref{difus}) splits into two identical equations for
$z<0$ and $z>0$. The solution is of the form
\begin{align}
n_{L}& =Ae^{-\omega _{+}z}+Be^{-\omega _{-}z}  \label{nlgral} \\
& -\left[ D\left( \omega _{+}-\omega _{-}\right) \right] ^{-1}\int_{0}^{z}
\left[ e^{-\omega _{-}\left( z-z^{\prime }\right) }-e^{-\omega _{+}\left(
z-z^{\prime }\right) }\right] \gamma \left( z^{\prime }\right) dz^{\prime },
\notag
\end{align}
where $\omega_\pm$, $A$ and $B$ are different on each side of the
wall. The parameters $\omega_\pm$ are given by
\begin{equation}
\omega _{\pm }=v_{w}/2D\pm \sqrt{\left( v_{w}/2D\right)
^{2}+\Gamma /D}, \label{omega}
\end{equation}
and the constants $A$ and $B$ are determined by the boundary
conditions $n_{L}\left( \pm \infty \right) =0$ and continuity of
$n_{L}$ and $n_{L}^{\prime }$ at $z=0$. Their general expressions
are rather cumbersome and can be found in the appendix. If the
$CP$ violating source $\gamma $ is localized on the wall, then
outside the wall $n_{L}\left( z\right) $ has a simple exponential
dependence [notice that, since $\omega _{+}>0$ and $\omega_{-}<0$,
only one exponential function survives on each side of the wall in
Eq. (\ref{nlgral})].

Since baryon number violation takes place in the
symmetric phase, we are only interested in the solution for $z>0$,
\begin{equation}
n_{L}\left( z\right) =Ae^{-\omega _{s+}z},  \label{nl}
\end{equation}
where $\omega_{s+}$ is given by Eq. (\ref{omega}), with
$\Gamma=a\Gamma_{ws}$, and the constant $A$ is an integral of
$\gamma \left( z\right) $ times a combination of exponentials. The
source $\gamma $ is proportional to $ v_{w}$, since it is given by
the flux of particles that reflect from the moving wall
\cite{ckn91}. Thus, $A$ is the product of $v_{w}$ times a factor
that depends on the details of $CP$ violation at the bubble wall.
Although this factor can be velocity-dependent, in general it is
not very sensitive to $v_{w}$. In the appendix we check this
assertion using an approximation for the source $\gamma \left(
z\right) $. According to these considerations, we will assume that
$ A\propto v_{w}$. Since we are interested in density contrasts
caused by a variation of the wall velocity, we only need to
determine the baryon density $n_{B}$ up to a constant factor
independent of $v_{w}$ (we assume that the $CP$ violation is
enough to generate the observed BAU).

The baryon number density also satisfies a diffusion equation,
similar to Eq. (\ref{difus}). In this case, only sphaleron
processes are relevant, so we have
\begin{equation}
D_{q}n_{B}^{\prime \prime }+v_{w}n_{B}^{\prime }-3\Gamma _{ws}\theta \left(
z\right) \left( n_{L}\left( z\right) +bn_{B}\right) =0.  \label{eqbaryon}
\end{equation}
Here, $D_{q}\sim 6T^{-1}$ is the diffusion coefficient for quarks, and the
numerical factor $b$ depends on the particle spectrum. The term proportional
to $n_{B}$ accounts for baryon number relaxation when the wall velocity is
small \cite{ck00,cmqsw01}. The chiral density $n_{L}$ acts as a source term
in Eq. (\ref{eqbaryon}) and  is given by Eq. (\ref{nl}).

The solution to Eq. (\ref{eqbaryon}) for $z\leq 0$ is a constant,
$n_{B}\equiv n_{B}\left( 0\right) $. For $z>0$, $n_{B}$ has the
form of Eq. (\ref{nlgral}), with $\gamma (z) =3\Gamma
_{ws}n_{L}(z)$ and
\begin{equation}
\omega _{B\pm }=v_{w}/2D_{q}\pm \sqrt{\left( v_{w}/2D_{q}\right)
^{2}+3b\Gamma _{ws}/D_{q}}.
\end{equation}
The conditions of continuity at $z=0$ and vanishing baryon number
at $ z=+\infty $ (i.e., far in the symmetric phase) give the value
of the constant baryon density in the broken symmetry phase,
\begin{equation}
n_{B}=\frac{3\Gamma _{ws}}{D_{q}\omega _{B+}} \int_{0}^{\infty
}n_{L}\left( z\right) e^{\omega _{B-}z}dz.  \label{nb}
\end{equation}
The parameter $\omega_{B-}$ is often approximated by
$\omega_{B-}\simeq -3b\Gamma_{ws}/v_w$, which is correct for
$v_w^2\gg 4 D_q\Gamma_{ws}$, i.e., for $v_w\gtrsim 10^{-2}$. Eq.
(\ref{nb}) can be easily integrated using Eq. (\ref{nl}),
\begin{equation}
n_{B}=\frac{Cv_{w}}{\left( v_{w}+\sqrt{v_{w}^{2}+ 4D_{q}3b\Gamma
_{ws}}\right) \left( v_{w}+\sqrt{v_{w}^{2}+4Da\Gamma _{ws}}\right)
+3bD\Gamma _{ws}}, \label{nbvw}
\end{equation}
where $C$ is a constant.

\begin{figure}[tbh] \centering
\epsfxsize=8cm \leavevmode \epsfbox{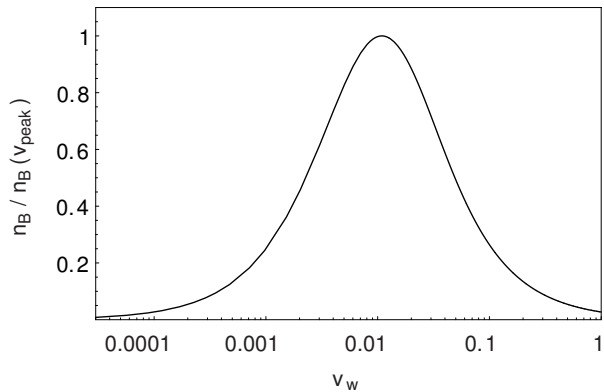} \caption{Baryon
number density as a function of the wall velocity} \label{fignbvw}
\end{figure}

The analytic approximation (\ref{nbvw}) for the dependance of the
BAU on the wall velocity is in reasonable agreement with other
approximations and numerical calculations
\cite{ck00,cmqsw01,ariel01,hn96}. Two quantitative differences
with some works arise due to our inclusion in Eq. (\ref{difus}) of
a non-vanishing rate $\Gamma\sim\Gamma_{ws}$ in the symmetric
phase, and the inclusion of the diffusion coefficient for quarks
in Eq. (\ref{eqbaryon}). In Fig.~\ref{fignbvw} we plot $
n_{B}\left( v_{w}\right) $ for $a= 3b= 1$. We see that the baryon
density has a peak, and tends to zero for small or large wall
velocities\footnote{Notice however that this approximation breaks
down for $v_w$ close to 1.}. For $D_q\ll D$, the peak is close to
$v_{w}= \sqrt{D\Gamma _{ws}}$. This can be seen either numerically
or by making approximations of Eq. (\ref{nbvw}) for the different
possible ranges of $v_w$. Furthermore, it can be easily seen that
Eq. (\ref{nbvw}) has the expected behavior on each side of the
peak, namely, $n_{B}\propto v_{w}$ for $v_{w}\ll \sqrt{D\Gamma
_{ws}}$, and $n_{B}\propto v_{w}^{-1} $ for $v_{w}\gg
\sqrt{D\Gamma _{ws}}$.

As we shall see, the wall velocity decreases during the phase
transition as a consequence of latent heat release. Therefore,
inside a given bubble, the local value of the baryon number
density at a distance $r $ from the center of nucleation is given
by $n_{B}(r)=n_{B}\left[ v_{w}(t)\right] $, where $t$ is  the
moment at which the bubble wall has passed through $r$, i.e., $t$
is given by $R\left( t\right)=r$. The total BAU is  the volume
average $ B=\int n_{B}\left[ v_{w}\left( t\right) \right] df$,
where $f\left( t\right) $ is the fraction of volume occupied by
bubbles. We remark that in this work we are assuming that $B$
gives the observed baryon abundance, which depends on the value of
the constant $C$ in Eq. (\ref{nbvw}). We are interested in the
relative variation of the baryon number density left behind by the
bubble walls, which does not depend on this constant, and can be
computed from Eq. (\ref{nbvw}) once the time dependance of the
wall velocity is known. Still, before closing this section we
would like to comment on the effects of the velocity decrease on
the total amount of baryons $B$.

If the initial velocity is $v_{i}\gg v_{\text{peak}}\sim
\sqrt{D\Gamma _{ws}} $, a decrease of $v_{w}$ produces an
enhancement of the BAU \cite{h95}. On the contrary, for $v_{i}\ll
v_{\text{peak}}$,  a velocity decrease will cause a suppression of
the generated baryon asymmetry \cite{ariel01}. It is interesting
to notice that in the case of an enhancement the effect can be
quite large; however, in the case of a suppression the electroweak
baryogenesis scenario will not be harmed significantly. Indeed,
the volume spanned by a bubble wall while the velocity is closer
to the initial value $v_i$  is in general of the same order of
that spanned with $v_w$ closer to the minimum velocity $v_m$.
Assuming that $n_B$ takes very different values $n_B\left(
v_i\right)$ and $n_B\left( v_m\right)$ in each of these volumes,
then the total baryon number density will be of the order of the
largest one. Even in models in which baryon densities with
opposite sign are formed in different regions, the suppression to
the total BAU should be of order one. Otherwise it would require a
fine tuning between the dynamics of the phase transition and the
baryogenesis mechanism.

\section{Phase transition and baryon inhomogeneities
\label{dynamics}}

The electroweak phase transition takes place in the
radiation-dominated era, at temperature $T_c\sim 100GeV$, when the
expansion rate is $H_{c}=1/2t_c$ and the age of the Universe is
given by
\begin{equation}
t_{c}=\xi M_{p}/T_{c}^{2}. \label{tc}
\end{equation}
Here, $M_p$ is the Planck Mass, and the parameter $\xi $  is
related to the light degrees of freedom of the plasma, $g_*$, by
$\xi =\sqrt{90/32\pi^3g_{\ast }}$ (at the electroweak scale,
$g_*\sim 100$ and $\xi \simeq 1/34$). The characteristics of the
transition are determined by the free energy density difference
$V\left( \phi ,T\right) $ between a value $\phi \neq 0$ of the
Higgs field VEV (corresponding to the broken-symmetry phase) and
$\phi =0$ (symmetric phase). In a first-order phase transition,
$V\left( \phi ,T\right) $ has a non-zero minimum $ \phi _{m}\left(
T\right) $ coexisting with the minimum $\phi =0$ and separated
from it by a barrier. Then, the quantity $V\left( T\right) \equiv
V\left( \phi _{m}\left( T\right) ,T\right) $ gives the pressure
difference between the two phases. In the simplest case, four
parameters characterize the behavior of $ V\left( \phi ,T\right)
$, namely, the critical temperature $T_{c}$ at which $ V\left(
T_{c}\right) =0$, the value of the order parameter at $T_{c}$, $
\phi _{m}\left( T_{c}\right)/T_c $, the latent heat (i.e., the
energy density discontinuity at $T=T_{c}$), given by
\begin{equation}
L=T_{c}\left( dV/dT\right) _{T_{c}}, \label{lat}
\end{equation}
and the tension of the bubble wall,
\begin{equation}
\sigma =\int \left( \frac{d\phi }{dr}\right) ^{2}dr,
\end{equation}
where $\phi \left( r\right) $ is the wall profile at $T=T_{c}$.

An important parameter, which is not directly related to the free
energy but is relevant for the dynamics of the phase transition is
the friction coefficient $\eta $. In a first-order phase
transition the system supercools, so bubbles  nucleate at a
temperature $T_{N}<T_{c}$. At $ T\leq T_{N}$ bubbles expand with a
velocity that is determined by the pressure difference between the
broken-symmetry phase inside the bubbles and the symmetric phase
that surrounds them, and by the friction of the wall with the
plasma \cite{lmt92,js01,dlhll92,mp95},
\begin{equation}
v_{w}\left( T\right) =-V\left( T\right) /\eta .  \label{velo}
\end{equation}

The electroweak wall velocity is subsonic, i.e. $v_w<c_s$, where
$c_s=1/\sqrt{3}$ is the velocity of sound in the relativistic
plasma. Therefore the bubble expands as a deflagration. The
deflagration front is the bubble wall, where the value of $\phi$
changes from $0$ to $\phi_m$ and the energy (\ref{lat}) is
released. A shock wave is formed in front of the wall, which is
preceded by a supersonic shock front \cite{k85}. The energy
released in the deflagration process is transmitted to the fluid
and distributed between the two fronts. Most of it is closer to
the phase transition front than to the shock front \cite{k85,h95},
so a temperature profile arises. However, if $v_w\ll c_s$ (which
is the case of the electroweak bubbles), part of this heat is
carried away far in front of the wall and can influence other
bubbles. If $v_w$ is small enough, one can assume that the
released latent heat has enough time to distribute uniformly and
equilibrate the temperature everywhere \cite{h95}. We will thus
assume for simplicity that the only effect of the shock wave is to
cause a uniform reheating of the plasma. This will be a good
approximation if the time it takes the latent heat to get
distributed throughout space (which depends on the bubble
separation) is much shorter than the time scales involved in the
development of the phase transition. We will examine the
consistency of this approximation in  section \ref{results}.

The phase transition thus develops essentially in two steps (see
e.g. \cite{ariel03}). The first stage is characterized by a quick
increase of temperature and a decrease of $v_{w}$, as the latent
heat reheats the plasma. After a certain time $\delta t_{1}$ the
plasma has reheated to a temperature that is close to the critical
one, and the system enters a phase-equilibrium or slow-combustion
stage \cite{w84,s82}. During this second stage the latent heat
release is compensated by the expansion of the Universe, and the
temperature remains almost constant. Since $T\simeq T_{c}$,
$V\left( T\right) \simeq 0$, so bubble expansion slows down
significantly, and bubble nucleation effectively stops. Hence,
this stage lasts for a longer time $\delta t_{2}$.

The time $\delta t_{\Gamma }$ during which bubbles nucleate is
much less than $\delta t_{1}$ and $\delta t_{2}$ \cite{ariel03}.
This is due to the quick variation of the nucleation rate $\Gamma
\left( T\right) $ with temperature. Therefore, it is a good
approximation to assume that all the bubbles nucleate at the
beginning of the phase transition (this is confirmed by our
numerical calculation). Hence, the number of bubbles remains fixed
during the phase transition, and their number density can only
change due to the dilution caused by the expansion of the
Universe. However, the time scale of the electroweak phase
transition is much less than the age of the Universe, so this
effect is negligible. This means that when the phase transition
completes, all the bubbles have roughly the same size, which is
$\sim n_b(T_N)^{-1/3}$.

The size scale of the inhomogeneities is given by the size of the
bubbles, and hence by the distance between centers of nucleation.
Their shape depends on the distances travelled by the bubble walls
during the times $\delta t_{1}$ and $\delta t_{2}$, and on the
baryon-number density generated during each stage. The amplitude
of the inhomogeneities (i.e., the contrast between the highest and
the lowest baryon number densities) is determined by the value
$v_{i}$ of the wall velocity at $ T\simeq T_{N}$ and by its
minimum value $v_{m}$ during the phase equilibrium stage.

Obtaining quantitative estimates of the phase transition from
analytical approximations is a difficult task. On one hand, the
dynamics involves integro-differential equations which complicate
the analysis. On the other hand, the exponential dependance of the
nucleation rate with temperature introduces a large amount of
uncertainty in the estimations. Nevertheless, an analytical
inspection will provide a qualitative picture of the
inhomogeneities and will be useful to study their parametric
dependance. We dedicate the rest of this section to such a study.
For that, we use the analytical expressions derived in Ref.
\cite{ariel03}. In the next sections we will check these results
by numerically integrating the equations for the progress of the
phase transition.

\subsection{Dynamics of the electroweak phase transition}

Bubbles begin to nucleate immediately after $T$ becomes smaller
than $T_c$. However, for $T$ close to $T_c$ the nucleation rate is
vanishingly small, so the number of bubbles is insignificant. The
onset of nucleation can be defined as the moment at which there
are enough bubbles so that they begin to feel the presence of each
other. More precisely, we define the temperature $T_N$ as that at
which the mean separation between bubbles equals the distance
travelled by sound waves (which carry latent heat) since time
$t_c$ \cite{ariel03}. Shortly after that, the temperature reaches
its minimum value $T_m$, say at time $t_m$, and then increases due
to the release of latent heat. Since the nucleation rate $\Gamma
(T)$ reaches its maximum at the minimum temperature $T_m$, and is
extremely sensitive to the temperature\footnote{For $T$ close to
$T_c$, the dependance of $\Gamma$ on $T$ is dominated by an
exponential function of the form $\Gamma\sim\exp [ -C/\left(
T_c-T\right)^2 ]$.}, it is sharply peaked at  $t=t_m$.

If the temperature $T_{N}$ is close enough to $T_{c}$, the thin
wall approximation can be used to estimate the nucleation rate.
In this case, the number density of bubbles is given by
\cite{ariel03}
\begin{equation}
n_{b}\sim \frac{2\pi \sigma ^{6}\eta ^{3}T_{c}^{2}}{9\xi
^{5}L^{8}}\left( \frac{T_{c}}{M_{P}}\right) ^{3}\left(
\frac{T_c}{T_{c}-T_{m}}\right) ^{9}T_{c}^{3},  \label{nbub}
\end{equation}
The difference $T_c-T_m$ can be roughly approximated
by\footnote{In this approximation we are missing a slight
(logarithmic) dependance of $T_c-T_m$ on the friction coefficient,
as can be deduced from Eq. (53) of Ref. \cite{ariel03}.}
$T_c-T_N$, where $T_N$ is given by
\begin{equation}
\left( \frac{T_{c}}{T_{c}-T_{N}}\right) ^{2}\simeq
\frac{3L^{2}T_{c}}{16\pi \sigma ^{3}} K,
\end{equation}
and
\begin{equation}
K=4\log \left( 2\xi M_{p}/T_{c}\right) +\log \left(
3L^{2}T_{c}/8\pi \sigma ^{3}\right) +6\log \left[ \left(
T_{c}-T_{N}\right) /T_{c}\right] \label{ka}
\end{equation}
is a dynamical factor related to bubble nucleation at $T\simeq
T_{N}$. This factor  depends on several parameters but its value
is dominated by the first term, so $K\sim 100$.

Notice that the number of bubbles depends on the friction $\eta$,
and thus on the wall velocity. This is because for large $v_w$
latent heat is released more quickly, so the plasma reheats faster
and the nucleation rate turns-off sooner.  The friction
coefficient  depends on the viscosity of the plasma and the
profile of the bubble wall. In the case $\phi _{m}\left( T\right)
\sim T$, which is required for baryogenesis, this dependence can
be factorized in the form \cite{ariel03}
\begin{equation}
\eta \simeq \tilde{\eta}T\sigma , \label{eta}
\end{equation}
where $\tilde{\eta}$ is a dimensionless constant that depends only
on the particle content of the plasma. The value of this parameter
determines the initial velocity $v_{i}$. Using the approximation
$V(T)\simeq V\left(T_N\right)\simeq L\left(T_N-T_c\right)$ in Eq.
(\ref{velo}), we can write the initial wall velocity in the form
\begin{equation}
v_{i}\simeq \frac{L\left(T_c- T_{N}\right) }{\eta T_c},
\label{vi}
\end{equation}
which makes apparent a factor of $v_i^{-3}$ in Eq. (\ref{nbub}).

As the Universe is reheated, bubble growth slows down. When
bubbles have transferred enough energy to the plasma, the
temperature reaches a value very close to $ T_{c}$ and stops
increasing. This happens if the latent heat $L$ is at least equal
to the difference $\delta \rho =\rho \left( T_{c}\right) -\rho
\left( T_{N}\right) $, where
\begin{equation}
\rho \left( T\right) =\pi ^{2}g_{\ast }T^{4}/30
\end{equation}
is the energy density of the plasma. If, on the contrary,
$L<\delta \rho $, then the temperature does not get close to
$T_{c}$ and the expansion does not slow-down significantly.

For $L>\delta \rho $, the critical temperature is reached when
the fraction of volume occupied by bubbles is
\begin{equation}
f_{1}\simeq \delta \rho /L.  \label{deltaf1}
\end{equation}
The average bubble radius at the end of the reheating stage is
thus
\begin{equation}
R_{1}\simeq \left( \frac{f_{1}}{4\pi n_{b}/3}\right) ^{1/3},
\label{r1}
\end{equation}
Notice that in Eq. (\ref{r1}) the radius $R_1$ does not have a
straight dependance on the value of the wall velocity during this
stage, as one could expect. This is because a larger $v_w$ implies
a shorter reheating time. In fact, the time $\delta t_{1}$ can be
estimated as $\delta t_{1}\simeq R_{1}/v_{i}$. However, the bubble
number density $n_b$ {\em does} depend on the initial value of
$v_w$, so in the end $R_1$ does depend on $v_i$ and $\delta t_1$
does not.

After reheating, bubbles can release latent heat only at the rate
at which this energy is taken away by the expansion of the
Universe. Therefore, the rate of bubble expansion is given by the
condition $L\dot{f}\simeq 4\rho H$, where $H= \sqrt{8\pi G\rho
/3}$ is the expansion rate (here, $G$ is Newton's constant).
Since the fraction of volume in the broken-symmetry phase is
$f\left( t\right) \simeq 4\pi n_{b}R\left( t\right) ^{3}/3$, we
can write $\dot{f}\simeq 4\pi n_{b} R_{2}^{2}v_{m}$, where the
mean radius $R_{2}$ during this stage is given by $n_{b}
R_{2}^{3}\sim 1$. Therefore, the wall velocity during slow
combustion is given by
\begin{equation}
v_{m}\sim \frac{\rho H}{Ln_{b}^{1/3}}.  \label{vf}
\end{equation}

One could naively think that the velocities (\ref{vi}) and
(\ref{vf}) are unrelated, since $v_{i}$ depends explicitly on the
friction $\eta $, whereas $v_{m}$ is determined by the Hubble rate
$H$. However, using the analytical approximations
(\ref{nbub}-\ref{ka}), the velocities in the two stages can be
written as
\begin{eqnarray}
v_{i} &\simeq &K^{-1/2}\frac{1}{\eta}\left( \frac{16 \pi
\sigma^{3}}{3 T}\right)^{1/2}
, \label{velocs}\\
v_{m} &\simeq &7g_{\ast }^{2/3}K^{-3/2}\frac{\sigma
^{5/2}T^{11/6}}{\eta L^{4/3}}.  \notag
\end{eqnarray}
We see that in this approximation $v_{m}$ is also proportional to
$\eta ^{-1} $, so the ratio $v_{i}/v_{f}$ does not depend on $\eta
$. This is a consequence of the fact that in Eq. (\ref{vf}), the
number density of bubbles is proportional to $\eta ^{3}$, whereas
$ \rho $ and $H$ are not directly related to $\eta $. Indeed, we
have seen that $n_b\propto v_{i}^{-3}$, so $v_{m}$ is in fact
proportional to $v_{i}$, the proportionality factor being
determined by the dynamics of reheating. From Eqs. (\ref{velocs})
we obtain
\begin{equation}
\frac{v_{i}}{v_{m}}\simeq\frac{4}{7}\frac{\left( L/T^{4}\right)
^{4/3}}{\sigma /T^{3}}\frac{K}{g_*^{2/3}}. \label{veloratio}
\end{equation}
Since $K$ and $g_*$ are both $\sim 100$, if $L$ and $\sigma $ were
of order 1 in temperature units, then $ v_{i}/v_{m}\lesssim 10$.
However, in the case of the electroweak phase transition we have
in general $L/T^{4}\lesssim 1$ and $ \sigma /T^{3}\ll 1$, so $
v_{i}/v_{m}$ can be quite large.

We remark that the initial velocity is determined by the viscosity
of the plasma and does not depend on the dynamics of the phase
transition, whereas the ratio $v_i/v_m$ depends only on the
dynamics of the phase transition, so $v_i$ and $v_i/v_m$ are quite
independent quantities.

A potentially important source of variation of $v_w$ is bubble
coalescence. When bubbles occupy a fraction of volume $ f\simeq
1/3$, they percolate. This means that at this point most bubbles
are in contact, so they can group to minimize surface energy. This
process contributes to the growth of bubbles and could dominate
the wall velocity \cite{w84}. When $f\simeq 1/2$ coalescence stops
because the regions of symmetric phase begin to form isolated
bubbles surrounded by the broken-symmetry phase, so the interfaces
are pushed again by the pressure difference between the two
phases. Hence, there may be a different velocity variation for
$1/3\lesssim f\lesssim 1/2$. To ascertain whether this is the case
we must compare the bubble growth rates caused by the two
mechanisms. The growth rate due to coalescence is $\dot{f}_{c}\sim
\left( \sigma n_{b}/\rho \right) ^{1/2}$ \cite{ariel03}. We
compare it with the rate due to pressure difference during the
slow combustion stage, $\dot{f}_{2}\sim \rho H/L$. Therefore we
have
\begin{equation}
\frac{\dot{f}_{c}}{\dot{f}_{2}}\sim \frac{L^{3/2}}{\sigma
T^{3}}\frac{K^{3/2} }{g_{\ast }^{3/4}v_{i}^{3/2}}\left(
\frac{T}{M_{p}}\right) ^{1/2}< 10^{-3}\frac{L^{3/2}}{\sigma T^{3}}
\end{equation}
for $v_i\gtrsim 10^{-2}$. For values of $L/T^{4}\sim 0.1-1$, it
would be necessary that $\sigma /T^{3}\ll 10^{-3}$ for coalescence
to dominate and produce a significant departure from the velocity
behavior described above. This does not seem likely for typical
values of the electroweak phase transition parameters, although in
general $\sigma/T^3\ll 1$.\footnote{In section \ref{results} we
consider values of $\sigma/T^3$ in the range $10^{-3}-10^{-1}$.
Notice that the values of $L$ and $\sigma$ are not unrelated since
both depend on the parameters of the theory. As a consequence,
lower values of $\sigma$ correspond in general to lower values of
$L$ \cite{ariel03}.}

\subsection{Generation of inhomogeneities}

The amplitude of the inhomogeneities is determined by the total
variation of the wall velocity during the transition. It is clear
that larger density contrasts will arise if the velocity variation
occurs far from the baryogenesis peak in Fig. \ref{fignbvw}, i.e.,
in the region where the baryon asymmetry is most sensitive to $
v_{w}$. If $v_{i}$ and $v_{m}$ are much larger than
$v_{\mathrm{peak}}$, we can use the approximation $n_{B}\propto
v_{w}^{-1}$. If, on the contrary, $ v_{i},v_{m}\ll
v_{\mathrm{peak}}$, then the baryon density has a dependance
$n_{B}\propto v_{w}$. In any of these two cases, the ratio of the
baryon densities in the high- and low- density regions,
$\epsilon=n_B^{\max}/n_B^{\min}$, is given by the ratio
$v_{i}/v_{m}$. In the intermediate case in which the variation of
$v_{w}$ crosses the peak of $n_{B}$, $\epsilon$ may be much
smaller.

In general the peak velocity is in the range $10^{-2}\lesssim
v_{\mathrm{peak} }\lesssim 10^{-1}$ \cite{ck00,cmqsw01}. Hence, to
give the largest possible amplitude, the wall velocity should be
either $v_{w }< 10^{-2}$ (in order to be $\ll 10^{-1}$) or $v_{w}>
10^{-1}$ (so that $v_w \gg 10^{-2}$). In general, however, the
electroweak bubble walls have an initial velocity $ v_{i}\sim
10^{-1}-10^{-2}$ \cite{js01,mp95}, so the velocity variation will
not occur too far from the peak of $n_B$. If the initial velocity
is $\sim 10^{-1}$, then $v_{w}$ will most likely cross
$v_{\mathrm{peak}}$ during the transition. So, the most favorable
situation is that in which $v_i$ is closer to $10^{-2}$. In this
case the velocity will not be far from $ v_{\mathrm{peak} }$
initially, but will probably not traverse it either. Large
inhomogeneities can then arise if the minimum velocity departs
enough from $v_i$.

As we have seen, the velocity variation depends on $L$ and $\sigma
$. According to Eq. (\ref{veloratio}), for large $L$ or small
$\sigma $, the wall velocity will change considerably, and
important baryon density contrasts may be formed. On the contrary,
if $L$ is too small or $\sigma $ too large, the wall velocity will
not change significantly during the transition. In this case, the
baryon number density will be rather homogeneous. Notice that for
$L\to 0$ there is no reheating at all, and we should have
$v_i/v_m\to 1$, so Eq. (\ref{veloratio}) fails in this limit. The
problem is that  for very small $L$, the bubble expansion never
enters the slow combustion stage, so the basic assumptions that
lead to Eq. (\ref{vf}) for the minimum velocity  break down.

Since all the bubbles nucleate in a short time $\delta t_\Gamma$
at the beginning of the transition, all the inhomogeneities should
have the same size scale, determined by the separation between
centers of nucleation, $d\sim n_b^{-1/3}$. As can be seen in Eqs.
(\ref{nbub}-\ref{eta}), $n_b$ is a very sensitive function of $L$,
$\sigma$, and the temperature difference $T_c-T_m$, which in turn
depends on $L$ and $\sigma$ too. Furthermore there is an
additional dependance on $\sigma$ through the friction coefficient
$\eta$. So, we cannot trust the dependance of $d$ on $L$ and
$\sigma$ obtained from these approximations. On the other hand, we
do not expect a significant dependance of $T_m$ on the friction,
so we can rely on the behavior $n_b\propto v_i^{-3}$, which
implies $d\propto v_i$.

\begin{figure}[tbh] \centering
\epsfxsize=8cm \leavevmode  \epsfbox{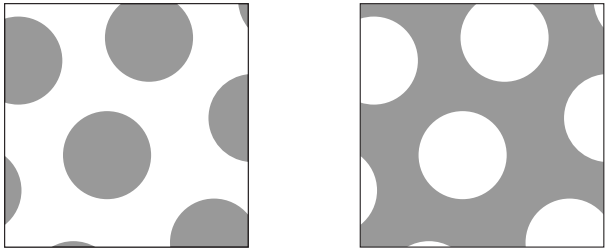}
\caption{Schematic picture of the geometry of the baryon
inhomogeneities created during bubble expansion. Shaded regions
correspond to high baryon number density. The left figure shows
the case in which $v_{i}<v_{\mathrm{peak}}$, whereas the right
figure shows the case $v_{m}>v_{\mathrm{peak}}$.} \label{geometry}
\end{figure}

The spherically symmetric inhomogeneities that originate inside
the bubbles can have two basic geometries, depending on the
relation between $v_{\mathrm{peak}}$, $v_i$, and $v_m$. If
$v_{i}<v_{\mathrm{peak}}$, the baryon asymmetry generated at the
bubble walls decreases with time as bubbles expand. Hence, a
larger baryon number density is produced at the centers of
bubbles, and then the geometry of the inhomogeneities is that of
spheres of high baryon number density $\sim n_B\left(v_i\right)$,
surrounded by regions with lower density $\sim
n_B\left(v_m\right)$ (see Fig. \ref{geometry}). These spheres have
radius $R_{1}$ and occupy a fraction of volume $f_1$, given by
Eqs. (\ref{deltaf1}) and (\ref{r1}).

In the case $v_{m}>v_{\mathrm{peak}}$, the baryon production will
be larger at the end of bubble expansion, and inhomogeneities will
have the form of walls which surround spherical voids of radius
$R_{1}$. Therefore, these walls have a width $w\sim
(f_1^{-1/3}-1)R_1$ and occupy a volume fraction $f_{2}=1- f_{1}$.
Of course, there is a third possibility, namely, that
$v_i>v_{\mathrm{peak}}$ and $v_m<v_{\mathrm{peak}}$, in which case
the structure of the inhomogeneities can be more complex. However,
as explained above, we do not expect inhomogeneities with large
amplitude when $v_w$ varies across $v_\mathrm{peak}$, so this case
is in fact irrelevant.

According to Eq. (\ref{deltaf1}), the volume fraction $f_1$ is
given by the ratio $\delta \rho /L$, which must be less than 1 in
order to have acceptable reheating and velocity variation. By
naturalness, $L$ is not likely to be too close to $\delta\rho$, so
$f_1$ is not expected to be very close to 1. This implies that the
width $w$ should be at least of the same order of $R_1$. On the
other hand, $f_1$ could be very small if $L\gg \delta \rho $. In
this case, we would have inhomogeneities of a small size
$R_{1}\sim \left( f_{1}/n_{b}\right) ^{1/3}$, separated by wide
walls of size $w\sim d\sim n_{b}^{-1/3}$. Notice however that
$R_1\ll d$ is not likely. Indeed, a size $R_1\sim 0.1 d$ already
requires a latent heat three orders of magnitude larger than the
energy difference between $T_N$ and $T_c$. Although $\delta
\rho/L$ can change significantly from one model to another
\cite{ariel03}, we do not expect that this ratio can be made
arbitrarily small in a realistic theory, since both the latent
heat and the amount of supercooling depend on the effective
potential. As a consequence, we expect that the sizes of the
spheres and walls in Fig. \ref{geometry} will be similar for
reasonable values of the parameters.  This is important because if
the size of the inhomogeneities is too small, they will be quickly
erased by diffusion and will not survive until later epochs.

\section{Numerical calculation of the the electroweak phase
transition \label{numerical}}

In order to simplify our numerical calculation of the phase
transition, we will use a free energy density of the form
\begin{equation}
V\left( \phi ,T\right) =D\left( T^{2}-T_{0}^{2}\right) \phi
^{2}-ET\phi ^{3}+ \frac{\lambda }{4}\phi ^{4}.  \label{veff}
\end{equation}
It is well known that to one-loop order in perturbation theory
and in the high-temperature approximation, the effective
potential takes this form, where the parameters $D$, $T_{0}$,
$E$, and $\lambda $ depend on the particle masses (see e.g.
\cite{quiros}). The actual form of the free energy depends on the
model and differs in general from (\ref{veff}). This may happen,
for instance, in the two-loop approximation, or in the case
perturbation theory or the high-temperature expansion are not
valid. In any case, this approximation is useful to simulate the
phase transition in different models, which is convenient for a
general treatment.

This is accomplished by adequately choosing the parameters in Eq.
(\ref{veff}) so that the free energy gives the correct values for
the thermodynamic parameters defined in the previous section. For
instance, the critical temperature and latent heat are given by
\begin{eqnarray}
T_{c} &=&T_{0}/\sqrt{1-E^{2}/\lambda D},  \notag \\
L &=&8D\left( E/\lambda \right) ^{2}T_{c}^{2}T_{0}^{2}.
\label{termoparam}
\end{eqnarray}
The first-order phase transition can take place between the
temperatures $ T_{c}$ and $T_{0}$, since in this temperature
range the potential (\ref{veff}) has a local minimum at $\phi
=0$, corresponding to the symmetric phase, and a global minimum at
\begin{equation}
\phi _{m}\left( T\right) =\frac{3ET}{2\lambda }\left[
1+\sqrt{1-\frac{8}{9} \frac{\lambda D}{E^{2}}\left(
1-\frac{T_{0}^{2}}{T^{2}}\right) }\right] ,
\end{equation}
corresponding to the broken-symmetry phase. These two minima are
separated by a barrier. At $T=T_{c}$ the minima are degenerate,
and at $T=T_{0}$ the barrier disappears and the point $\phi=0$
becomes a maximum. The pressure difference between the two phases
is $V\left( T\right) \equiv V\left( \phi _{m}\left( T\right)
,T\right) $. The order parameter of the phase transition is given
by
\begin{equation}
\phi _{c}/T_c\equiv \phi _{m}\left( T_{c}\right)/T_c =2E/\lambda .
\label{fic}
\end{equation}

The progress of the transition is characterized by the fraction of
volume occupied by bubbles of the broken-symmetry phase. In the
case of the electroweak phase transition, this is given by
\cite{gw81,ah92,dlhll92}
\begin{equation}
f\left( t\right) =1-\exp \left\{ -\frac{4\pi }{3}\int_{t_{c}}^{t}\Gamma
\left( T^{\prime }\right) R\left( t^{\prime },t\right) ^{3}dt^{\prime
}\right\} .  \label{fraction}
\end{equation}
Here, $R\left( t^{\prime },t\right) $ is the radius of a bubble
that nucleated at time $t^{\prime }$ (and temperature $T^{\prime
}$) and expanded until time $t$,
\begin{equation}
R\left( t^{\prime },t\right) \simeq \int_{t^{\prime }}^{t}v_{w}\left(
T^{\prime \prime }\right) dt^{\prime \prime },  \label{radius}
\end{equation}
where the wall velocity is given by Eq. (\ref{velo}), and we have neglected
the initial radius of the critical bubble. The nucleation rate $\Gamma
\left( T\right) $ is given by \cite{a81,l83}
\begin{equation}
\Gamma \simeq T_{c}^{4}e^{-S_{3}/T},  \label{gamma}
\end{equation}
where $S_{3}\left( T\right) $ is the three-dimensional instanton
action\footnote{The prefactor of the exponential in (\ref{gamma})
is roughly assumed to be of order $T^{4}$, since the rate is
dominated by the exponential.}, which coincides with the free
energy of a critical bubble in unstable equilibrium between
expansion and contraction,
\begin{equation}
S_{3}=4\pi \int_{0}^{\infty }r^{2}dr\left[ \frac{1}{2}\left(
\frac{d\phi }{dr}\right) ^{2}+V\left( \phi \left( r\right)
,T\right) \right] .
\end{equation}

The configuration of the nucleated bubble can be obtained by
extremizing this action. Hence it obeys the equation
\begin{equation}
\frac{d^{2}\phi }{dr^{2}}+\frac{2}{r}\frac{d\phi
}{dr}=\frac{\partial V}{\partial \phi }.  \label{eqprofile}
\end{equation}
For $T\to T_c$ the radius of the bubble becomes infinite, so the
profile of the wall can be easily calculated by neglecting the
second term in (\ref{eqprofile}). Then, we can readily compute the
wall tension for the model (\ref{veff}),
\begin{equation}
\sigma =\frac{2\sqrt{2}E^{3}}{3\lambda ^{5/2}}T_{c}^{3}.
\end{equation}
This equation, together with Eqs. (\ref{termoparam}) and
(\ref{fic}) determine completely the values of the parameters of
$V\left( \phi ,T\right) $ in terms of thermodynamical properties
of the free energy.

Analytical approximations for the profile of the critical bubble
introduce large errors in the nucleation rate. Therefore, we will
use the numerical fit for $S_{3}$ given in Ref. \cite{dlhll92},
\begin{eqnarray}
\frac{S_{3}}{T} &=&13.72\frac{E}{\lambda ^{3/2}}\ \alpha
^{3/2}g\left(
\alpha \right) ,  \label{s3} \\
g\left( \alpha \right) &=&1+\frac{\alpha }{4}\left(
1+\frac{2.4}{1-\alpha }+ \frac{0.26}{\left( 1-\alpha \right)
^{2}}\right) ,  \notag
\end{eqnarray}
where $\alpha =\lambda D\left( T^{2}-T_{0}^{2}\right)
/E^{2}T^{2}$. This dimensionless parameter can also be written as
\begin{equation}
\alpha
=\frac{T_{c}^{2}}{T_{0}^{2}}\frac{T^{2}-T_{0}^{2}}{T_{c}^{2}-T_{0}^{2}
},  \label{alfa}
\end{equation}
where it is apparent that it varies from $1$ to $0$ as $T$
decreases from $T_{c}$ to $T_{0}$.

Finally, the variation of temperature with time during the phase
transition is given by \cite{ariel03}
\begin{equation}
T^{3}=\frac{T_{c}^{3}a_{c}^{3}}{a^{3}}+\frac{V^{\prime }\left(
T\right) }{ 2\pi ^{2}g_{\ast }/45}f.  \label{temperature}
\end{equation}
Here, the prime means total derivative with respect to
temperature. The first term in Eq (\ref{temperature}) gives the
usual variation of $T$ in the adiabatic expansion, while the
second term accounts for the release of entropy during the phase
transition. The evolution of the scale factor $a\left( t\right) $
is given by the Friedman equation, $\left( \dot{a}/a\right)
^{2}\equiv H^{2}=8\pi G\rho \left( T\right) /3$. Since the
duration of the electroweak phase transition is much shorter than
the age of the Universe, we can use the approximation $H\simeq
H_{c}$ \cite{ariel03}. Note that in Eqs. (\ref{fraction}) and
(\ref{radius}) we have neglected the variation of length scales
due to the expansion of the Universe. We cannot do the same in
Eq. (\ref{temperature}), since for $ T\simeq T_{c}$ small changes
of temperature are relevant. Indeed, the scale of temperature
variation in a first order electroweak phase transition is given
by $ T_{c}-T_{0}\simeq \left( E^{2}/2\lambda D\right) T_{c}\ll
T_{c}$.

\section{Numerical results} \label{results}

In order to solve numerically Eqs.
(\ref{fraction}-\ref{temperature}) it is convenient to
differentiate Eq. (\ref{temperature}). At this stage we can make
the approximations $a\simeq a_{c}$, $T\simeq T_{c}$, so we have
\begin{equation}
\left( 1-\frac{15}{2\pi ^{2}g_{\ast }}\frac{V^{\prime \prime
}}{T^{2}} f\right)
\frac{dT}{T_{c}}=-\frac{1}{2}\frac{dt}{t_{c}}+\frac{15}{2\pi
^{2}g_{\ast }}\frac{V^{\prime }}{T^{3}}df.  \label{developt}
\end{equation}
In addition, the variation of $f$ is given by
\begin{equation}
df=\left( 1-f\right) 4\pi v_{w}\left( t\right)\left[
\int_{t_{c}}^{t}\Gamma R^{2}dt^{\prime }\right] dt.
\label{developf}
\end{equation}
Integration of these equations gives the values of the radius
$R(t)$ and wall velocity $v_w(t)$ at every time. Then, the profile
$n_B(r)$ is obtained with the aid of Eq. (\ref{nbvw}), as
explained at the end of section \ref{baryo}. For the numerical
calculation we have made a simple discretization of Eqs.
(\ref{developt}) and (\ref{developf}). We have used a time step
much shorter than the duration of the phase transition ($\sim
1/1000$), and we have verified that the result remained unchanged
as this step decreased further. Therefore, the uncertainties of
the numerical calculation are certainly smaller than the ones
introduced by the analytical approximations used in the previous
section.

\subsection{Parameter values}

Several extensions of the Standard Model provide a strong phase
transition as well as  enough CP violation for baryogenesis.
Supersymmetric models were extensively studied in the literature,
especially the case of the MSSM, which gives the required BAU in a
light right-handed stop scenario
\cite{ck00,cmqsw01,hn96,cqw96,lr01}. Some alternatives to this
model include non-minimal supersymmetric extensions (see e.g.
\cite{p93}), the addition of  heavy fermions to the SM
\cite{cmqw04}, a dimension-six operator in the effective Higgs
potential \cite{gsw04}, or the presence of hypermagnetic fields
during the phase transition \cite{e98,abpp01}. Therefore, we will
consider various values of the parameters.

For the case of the MSSM, the thermodynamic parameters of the
phase transition have been calculated \cite{lr01}. Hence, we will
take this case as a starting point for the parameter variation. We
will consider different values of $L$ and $\sigma$, since these
two parameters are the ones which most directly affect the
dynamics of the phase transition. Thus, according to the
non-perturbative study of Ref. \cite{lr01}, our reference values
will be $T_{c}\simeq 85GeV$, $\phi _{c}/T_{c}\simeq 1$, $
L/T_{c}^{4} \simeq 0.4$, $\sigma /T_{c}^{3}=0.01$. The viscosity
of the plasma has also been studied for the MSSM in the light stop
scenario \cite{js01}. We will consider values of the friction
parameter $ \tilde{\eta}\sim 0.1-10$, which give initial
velocities\footnote{We obtained these relations between the values
of $\tilde{\eta}$ and $v_i$ numerically. The values of $v_i$
obtained from Eqs. (\ref{velocs}) and (\ref{eta}) are a factor
$\simeq 2.5$ less than the numerical values.} $v_{i}\sim 10^{-1}-
10^{-2}$, in accordance with the results of Ref.\cite{js01}.

To illustrate the kind of profiles of the baryon number density
that can be created inside an expanding bubble, we will consider
several values of the electroweak baryogenesis parameters.
According to the discussions in section \ref{dynamics}, the
profile depends essentially on the position of $v_{\rm peak}$
relative to the velocity variation. We have seen in section
\ref{baryo} that  setting the values of the parameters $a$ and $b$
to $\sim 1$, the peak velocity is given by $v_{\rm peak}\sim
\sqrt{D\Gamma_{ws}}$. For the MSSM, $D\sim 100T^{-1}$, so this
gives the correct value $v_{\rm peak}\sim 10^{-2}$. In other
extensions of the SM the effective diffusion constant $D$ may
differ from that value, and further deviations may arise through
the precise values of $a$ and $b$. Therefore, we will consider
values of $v_{\rm peak}$ in the range $10^{-3}-10^{-1}$.

\subsection{Dynamics of the phase transition}

\begin{figure}[tbh] \centering \epsfxsize=8cm
\leavevmode  \epsfbox{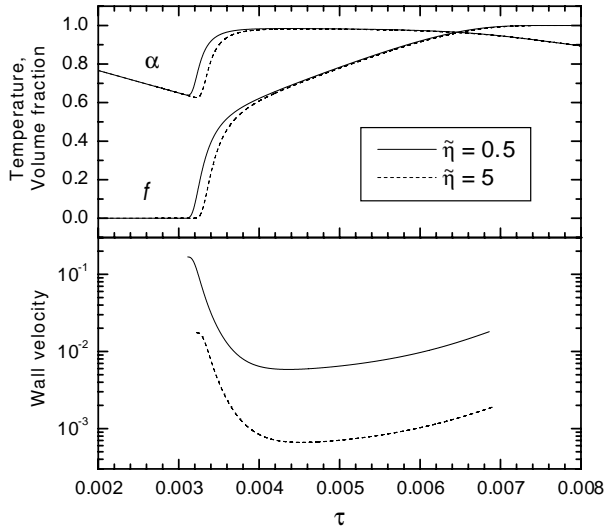} \caption{Fraction of volume,
temperature parameter $\protect\alpha $, and wall velocity as
functions of dimensionless time $\protect\tau =\left(
t-t_{c}\right) /t_{c}$.} \label{figdevelop}
\end{figure}

The evolution of the phase transition for these values of the
parameters is shown in Fig. \ref{figdevelop}. We have plotted the
variable $\alpha $ defined in Eq. (\ref{alfa}), the fraction of
volume $f$ occupied by bubbles, and the bubble wall velocity
$v_{w}$, as functions of time. Notice that the evolution of the
temperature and the fraction of volume is affected by the bubble
wall friction only at the beginning of the phase transition. This
is because during the reheating stage, the bubble expansion is
dominated by friction. On the contrary, during the phase
equilibrium stage the development of the phase transition is
determined by the balance between the rate of latent heat release
and that of cooling due to the expansion of the Universe, as
explained in section \ref{dynamics}.

We see that the wall velocity decreases by a factor that does not
depend on the friction coefficient (and hence on the initial
velocity $v_{i}$), as anticipated by the analytical
approximations. In the present case the factor is $\simeq 30$, to
be contrasted with the estimation (\ref{veloratio}), which for the
parameters under consideration gives  $v_i/v_m \simeq 100$. As
expected, the discrepancy is of order 1.

Notice also that the wall velocity is not constant in the slow
combustion stage, but it slowly begins to grow again after
reaching its minimum value $v_{m}$. This happens because as the
regions occupied by the symmetric phase become smaller, the energy
must be released more quickly at the interfaces in order to
compensate the cooling produced by the adiabatic expansion.

\begin{figure}[tbh]
\centering \epsfxsize=8cm \leavevmode  \epsfbox{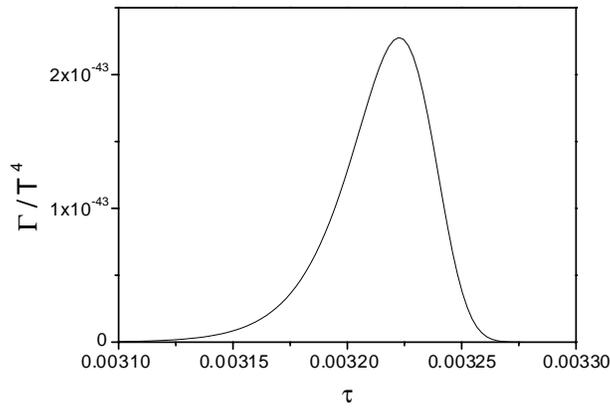}
\caption{The nucleation rate as a function of dimensionless time
$\tau$ defined in Fig. \ref{figdevelop}.} \label{figgamma}
\end{figure}

In the analysis of the inhomogeneities we have assumed that all
bubbles begin to expand at the time $t_i$ at  which the nucleation
rate $\Gamma$ takes its maximum value. We have checked numerically
that most of the bubbles nucleate at the very beginning, while the
volume fraction increases from $10^{-3}$ to $10^{-2}$. Moreover,
Fig. \ref{figgamma} shows that $\Gamma$ is active during a time
interval $\delta t_{\Gamma }$ which is two orders of magnitude
shorter than the total duration of the phase transition (compare
with Fig. \ref{figdevelop}). This introduces a dispersion of about
a 5\% in the bubble radius at the end of the transition, which we
define to be the moment $t_f$ at which $ f=0.99$ (the initial and
final times delimit the plot of the wall velocity in Fig.
\ref{figdevelop}). Hence, there will not be inhomogeneities at
different scales, although some dispersion in sizes will arise due
to the fact that the centers of nucleation are randomly scattered
throughout space.

\subsection{Profile of the baryon inhomogeneities}

\begin{figure}[tbh]
\centering \epsfxsize=8cm \leavevmode  \epsfbox{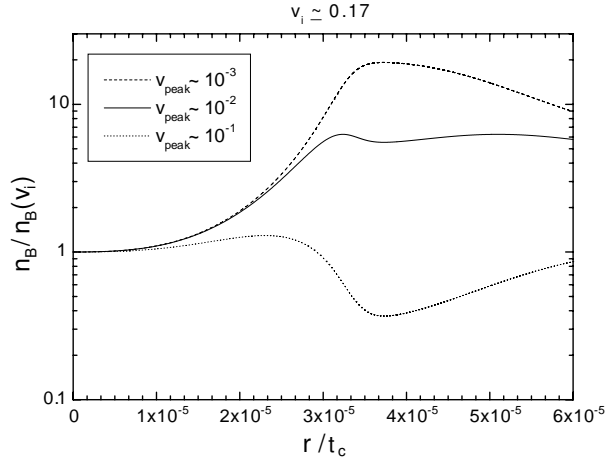}
\caption{Baryon number density as a function of the bubble radius
$r$ in units of the age of the Universe $t_{c}$, for $v_{i}\sim
0.1$.} \label{figprofile1}
\end{figure}

\begin{figure}[tbh]
\centering \epsfxsize=8cm \leavevmode  \epsfbox{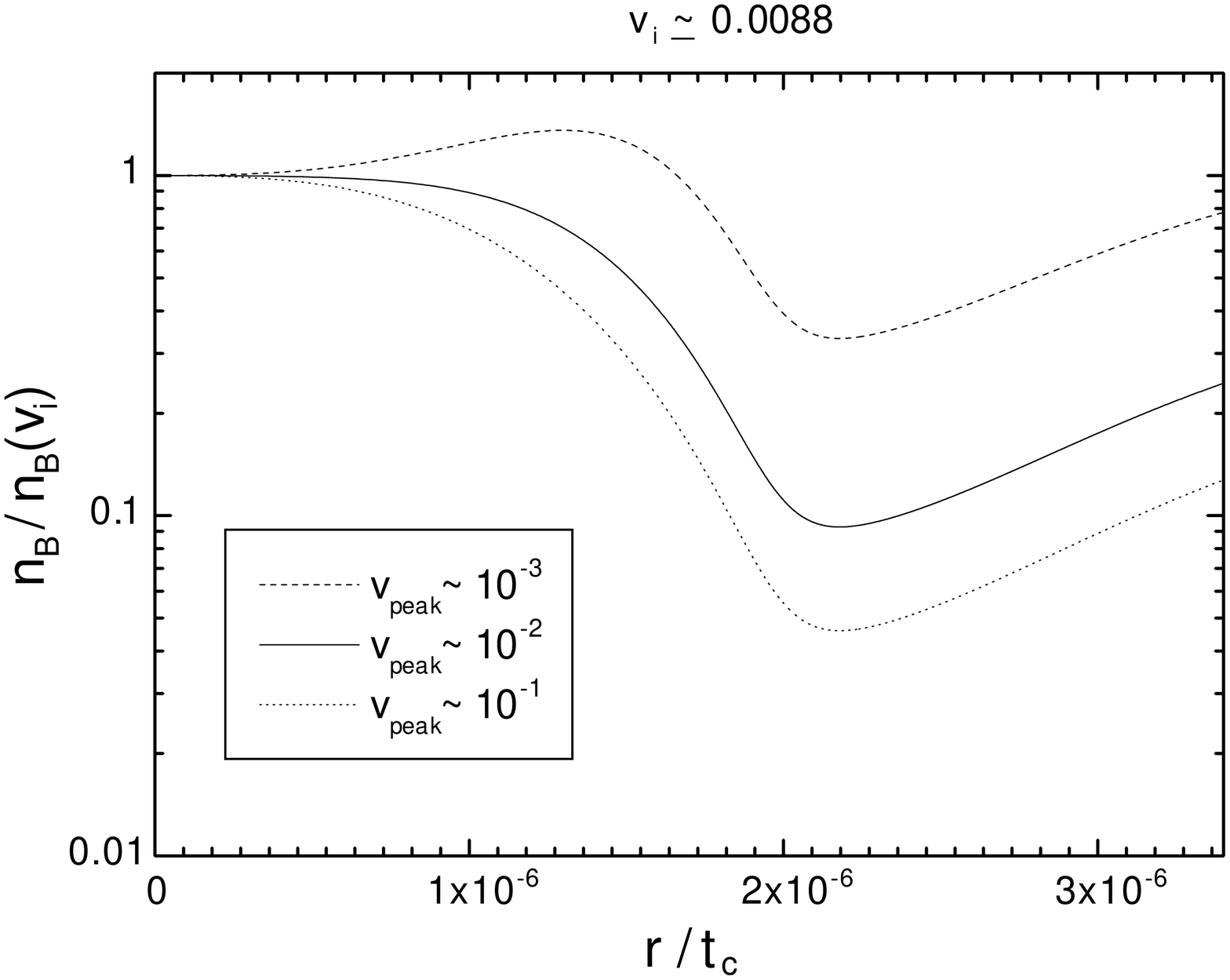}
\caption{Baryon number density as a function of the bubble radius
$r$ in units of the age of the Universe $t_{c}$, for $v_{i}\sim
0.01$.} \label{figprofile2}
\end{figure}

In Figs. \ref{figprofile1} and \ref{figprofile2} we show the
inhomogeneity profiles that arise for initial wall velocities
slightly above $0.1$ and slightly below $0.01$ respectively, and
different possible values of $v_{\mathrm{peak}}$.

In the case $ v_{i}\gtrsim 0.1$, we see that larger amplitudes are
obtained for smaller values of $v_{\mathrm{peak}}$, i.e., when
$v_{i}$ is more distant from the baryon generation maximum. The
profile for $v_{\mathrm{peak}}\sim 10^{-3}$ in Fig.
\ref{figprofile1} is in good agreement with the result of Ref.
\cite{h95}, where a dependance $n_B\propto v_w^{-1}$ was assumed.
The maximum of this curve corresponds to the minimum wall velocity
$v_{m}$ reached during the phase transition, since in this case
the velocity $v_{\mathrm{peak}}$ is not attained. In the other
curves, instead, the peak in $n_B(v_w)$ becomes manifest. In the
case $v_{\mathrm{peak}}\sim 10^{-2}$, the amplitude is smaller,
and we observe the presence of two hills in the curve. This is
because $v_{w}$ crosses $v_{\mathrm{peak}}$ immediately before
reaching the minimum $v_m$. Then it crosses the peak again due to
the final acceleration. The case $v_{\mathrm{peak}}\sim 10^{-1}$
gives still a smaller amplitude, since the velocity variation is
more symmetric around the peak.

Fig. \ref{figprofile2} shows the case $v_{i}\lesssim 10^{-2}$. As
expected from the discussion of section \ref{dynamics}, the
profiles behave quite oppositely to the previous case. The largest
amplitude is obtained for $v_{\mathrm{peak}}\sim 10^{-1}$, and in
this case the baryon number density is higher at the centers of
the bubbles. The upper curve is similar to the lower curve of Fig.
\ref{figprofile1}, because in this case the wall velocity crosses
the peak. We see that when this happens, the amplitude is smaller
and the baryon inhomogeneities are not clearly localized either in
the interior or the exterior of the spheres of Fig \ref{geometry}.

It is important to notice that these profiles describe the
inhomogeneities produced by an average bubble that expands without
colliding with other bubbles. Furthermore, bubble collisions may
have some effect on the development of the phase transition. Since
Eq. (\ref{fraction}) takes into account overlapping of bubbles,
the present simulation of the phase transition is in principle
reliable in the whole time interval, with the possible exception
of collisions and coalescence, which occur when the fraction of
volume is $f\sim 0.3-0.5$. We have seen in section \ref{dynamics}
that the dynamics of the electroweak phase transition is not
affected by bubble coalescence. However, we are neglecting some
possible effects, such as additional reheating due to bubble
collisions. Such effects could introduce some distortion in the
profiles shown in Figs. \ref{figprofile1} and \ref{figprofile2}.

\subsection{Size and amplitude of the inhomogeneities}

\begin{figure}[tbh]
\centering \epsfxsize=8cm \leavevmode  \epsfbox{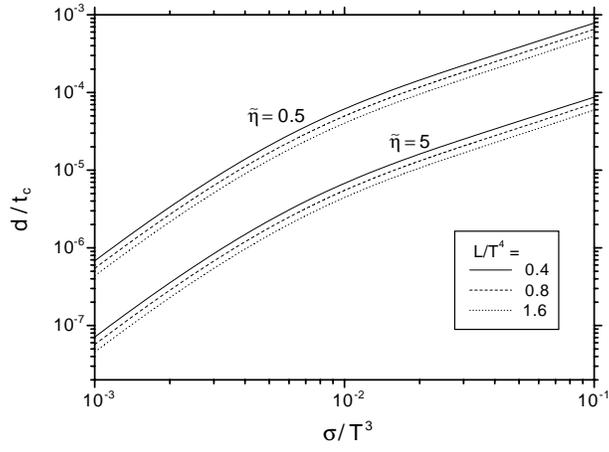}
\caption{Size of the inhomogeneities as a function of the bubble
wall tension for different values of the latent heat and
friction.} \label{figsize}
\end{figure}

The size scale of the inhomogeneities is given by the distance
between centers of bubbles, $d\sim n_{b}^{-1/3}$ \cite{h95}. We
plot this distance in Fig. \ref{figsize} as a function of the
bubble wall tension for different values of the latent heat and
friction. Notice that, as expected, changing  $\eta$ by an order
of magnitude induces the same variation in $d$. We have studied
the dependance of $d$ on the initial wall velocity in the range $
10^{-3}< v_i < 0.4$, and we have checked that the linearity
$d\propto v_i$ is verified for the different values of $L$ and
$\sigma$.

\begin{figure}[tbh]
\centering \epsfxsize=8cm \leavevmode  \epsfbox{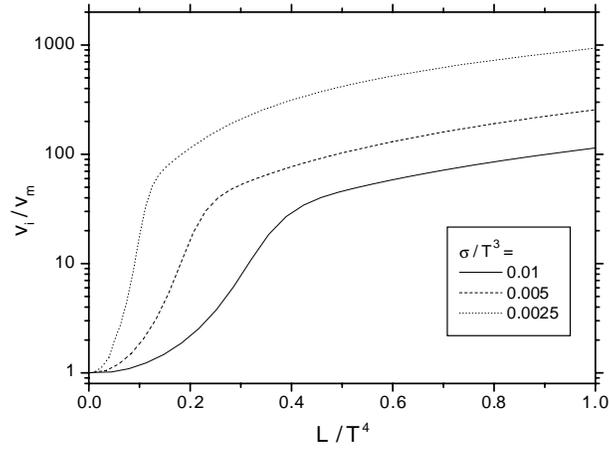}
\caption{The velocity ratio $v_i/v_m$ as a function of latent heat
for different values of the bubble wall tension.} \label{figvivm}
\end{figure}

The maximum possible baryon density contrast $\epsilon$ is
determined by the ratio $v_i/v_m$. In Fig. \ref{figvivm} we plot
this ratio as a function of latent heat, for different values of
$\sigma$. For MSSM parameters we find $v_i/v_m\sim 10-100$. We
observe that for large $L$, Eq. (\ref{veloratio}) gives the
correct order of magnitude and the exact parametric dependance. In
contrast, for smaller values of latent heat, the behavior of the
numerical curves changes abruptly. This was expected, since when
$L\lesssim \delta \rho$ the reheating becomes insufficient and the
phase transition ceases to have two well defined stages. Thus, the
curves depart from the approximate behavior (\ref{veloratio}), and
for $L\to 0$, $v_m$ approaches $v_i$, as required physically.

In this work we have assumed that the latent heat released as  the
bubbles expand is quickly distributed, so that the only effect is
a global reheating. To estimate whether this approximation is
correct or not, we must compare the time scales for the evolution
of the phase transition with the time in which the latent heat
liberated in a deflagration front reaches the neighboring bubbles.
As seen in Fig. \ref{figdevelop}, for our reference values of the
parameters  the quick reheating stage occurs in a time $\delta
t_1\sim 10^{-4} t_c$. On the other hand, the time it takes a shock
front to travel the distance between bubbles is given by $\delta
t_{sh}\sim d$. Hence, we expect that our treatment will be
reliable for $d/t_c\ll 10^{-4}$. According to Fig. \ref{figsize},
for  $v_i \gtrsim 10^{-1}$ ($\tilde{\eta}=0.5$) this definitely
requires $\sigma/T^3<10^{-2}$. For smaller values of the initial
wall velocity this condition is relaxed. After the system reheats,
the transition proceeds more slowly. In this stage the time scale
is $\delta t_2\sim 10^{-3}t_c$, and the approximation breaks down
only for higher values of the wall velocity and tension, $v_i\sim
0.1$, $\sigma/T^3 \sim 10^{-1}$.  Since the amplitude of the
inhomogeneities is larger for smaller values of $\sigma$ (see Fig.
\ref{figvivm}), in the interesting cases the global reheating
approximation is valid.

\section{Effect on the quark-hadron phase transition} \label{conseq}

The aim of this paper was to study the characteristics of the
baryon inhomogeneities generated in the electroweak phase
transition. However, we wish to comment briefly on the
possibilities for these inhomogeneities to leave a sequel in the
subsequent evolution of the Universe. The time scale for the
wash-out of the inhomogeneities depends on their size and
amplitude, and on the diffusive processes. For temperatures in the
range $100GeV<T<1MeV$, the only important dissipation process is
neutrino diffusion. A wide range of sizes and amplitudes is left
almost unaffected by this process \cite{jf94,hh93}. Therefore, it
is likely that electroweak inhomogeneities survive somewhat
unchanged until the QCD scale \cite{s03,jf94}, in which case they
may influence the dynamics of the quark-hadron phase transition.
Below $1MeV$ other processes (namely, baryon and photon diffusion)
become important and may completely erase the electroweak
inhomogeneities before the nucleosynthesis epoch.

The evolution of baryon inhomogeneities between the epochs of
$T=100GeV$ and $T=100MeV$ has been considered in Ref. \cite{jf94}.
According to this calculation, a baryon inhomogeneity which at the
electroweak scale has amplitude $<10^3$ and size $>10^{-8}t_c$,
corresponding to the cases considered in Figs. \ref{figsize} and
\ref{figvivm}, may survive almost unchanged until the QCD epoch.

As shown in Fig. \ref{figvivm}, in the MSSM inhomogeneities with
amplitude $\epsilon_\textrm{MSSM}\sim 10-100$ are likely. However,
there exist several interesting scenarios  for the electroweak
phase transition \cite{p93,cmqw04,gsw04,e98,abpp01}  besides the
MSSM. In some of them the phase transition is much stronger than
in the latter. In such extensions of the SM, the parameters
affecting the dynamics will differ significantly from those of the
MSSM. In particular, larger values of the latent heat are expected
in stronger phase transitions \cite{ariel03}. This leads in
general to a larger $\epsilon$,  as can be seen in Fig.
\ref{figvivm}. Further enhancement (or suppression) may also
result from the variation of the wall tension. In fact, one
expects larger values of $\sigma$ in stronger phase transitions,
so in principle this effect goes in the opposite direction to that
of latent heat. In any case, a sensible estimation of the
inhomogeneity amplitude requires a calculation of $L$ and $\sigma$
in each particular case. Nevertheless, it is reasonable to expect
that in some models amplitudes considerably larger than in the
MSSM may arise. In such models we will possibly find $\epsilon\gg
100$.

As can be seen in Figs. \ref{figsize} and \ref{figvivm}, values of
$L$ and $\sigma$ which give larger amplitudes, yield also smaller
sizes. Such a correlation is important because inhomogeneities
with larger amplitudes and smaller sizes are more easily diluted.
Extrapolating our results beyond the ranges of parameters
considered in the previous section, we see that for an amplitude
$\sim 10^4$ we would have a size $d/t_c\sim 10^{-8}-10^{-7}$,
which is still undamped by neutrino inflation. Larger amplitudes
will probably be affected. For instance, according to the results
of Ref. \cite{jf94}, an initial amplitude $\sim 10^5$ may decrease
to $\sim 10^4$ if the size scale is $\lesssim 10^{-9}t_c$.
Therefore, it seems unlikely that inhomogeneities generated in the
electroweak phase transition may have an amplitude much larger
than $10^4$ at the time of the QCD phase transition.

In Ref. \cite{s03}, it was found that a significant effect on the
latter is obtained for inhomogeneities with an amplitude
$\epsilon\sim 10^7$. However, it was also noticed that smaller
values of $\epsilon$ will have similar effects if the amount of
supercooling is smaller than the one considered there, in
accordance to Ref. \cite{bg92}. Furthermore, in the treatment of
Ref. \cite{s03} the possibility that the nucleation rate changes
with the chemical potential (see e.g. \cite{ck92}) was not
considered. Taking into account this effect could affect strongly
the discussion in Ref. \cite{s03}, and therefore lead to a
significantly different value of the inhomogeneity amplitude
needed to affect this transition. Altogether, it is clear that the
dynamics of the quark-hadron phase transition is still not fully
understood, and we can sensibly expect that it may be affected by
prior inhomogeneities with amplitudes of the order of those that
are generated in the electroweak phase transition. We hope to
address this issue in future work.

\section{Conclusions} \label{conclu}

In this paper we have made a detailed study of the baryon number
inhomogeneities that can be generated as a byproduct of
electroweak baryogenesis. In particular, we have investigated
analytically the parametric dependance of the inhomogeneities, and
we have made a more precise calculation of the amplitude and size
of the density contrasts by solving numerically the equations for
the development of the electroweak phase transition. By doing
that, we have checked several analytical approximations derived in
Ref. \cite{ariel03} for the dynamics of the phase transition. An
important difference between our treatment and previous analysis
is that we have taken into account the fact that in general the
baryon density has a maximum at a bubble wall velocity
$v_w=v_\mathrm{peak}$. We have accomplished this by deriving a
simple analytical approximation for the dependance of the baryon
asymmetry of the Universe on $v_w$.

The characteristics of the inhomogeneities depend on the
parameters that determine the dynamics of the phase transition,
namely, the latent heat $L$, bubble wall tension $\sigma$, and
friction coefficient $\eta$, and on the parameters that govern the
production of baryons, which are essentially the diffusion
constant $D$ for the chiral quark asymmetry and the weak sphaleron
rate $\Gamma_{ws}$. Some general features, however, are
independent of these parameters. For instance, the spherical
symmetry and the absence of a significant dispersion in sizes.

Indeed, we have shown that all the bubbles are nucleated in a
short time interval at the beginning of the transition, and
therefore have roughly the same radius. (We wish to remark here
that this feature of first-order phase transitions may be
important for other cosmological consequences as well, such as the
formation of topological defects or magnetic fields. We will
address these issues somewhere else.) Since the baryon
inhomogeneities are generated in the walls of expanding bubbles
and depend on the wall velocity, this implies that all
inhomogeneities have approximately the same size, amplitude, and
profile.

The size scale is given by the distance between centers of
nucleation, so it depends only on the phase transition parameters
$L$, $\sigma$, and $\eta$. In particular, we have seen that the
higher the initial wall velocity, the larger the bubble
separation, due to the sooner turn-off of the nucleation rate. The
amplitude $\epsilon$ of the density contrasts may depend further
on the baryogenesis parameters; more precisely, on
$v_\mathrm{peak} \simeq\sqrt{D\Gamma_{ws}}$. However, if  $v_w$
stays far from $v_\mathrm{peak}$ throughout the phase transition,
$\epsilon$ is given by the ratio $v_i/v_m$ between the maximum and
minimum values of $v_w$. We have seen that this ratio does not
depend on the friction parameter, so in this case $\epsilon$ is
only determined by $L$ and $\sigma$. If the wall velocity crosses
the value $v_\mathrm{peak}$ during the transition, then the
amplitude is smaller than $v_i/v_m$.

The exact profile of the inhomogeneities, on the other hand,
depends both on $v_\mathrm{peak}$ and on the dynamics of the phase
transition. In the case of the MSSM, the  initial velocity is most
likely close to the baryogenesis peak, $v_i\sim
v_\mathrm{peak}\sim 10^{-2}$ \cite{cmqsw01,js01}. As we have seen,
this implies that the geometry is that of high-density spheres
surrounded by low-density walls. This is contrary to the case of
the SM, where wall-shaped inhomogeneities are formed \cite{h95},
showing that both kinds of profile can arise in different models.

For MSSM values of the parameters $L$ and $\sigma$, we find
amplitudes $\epsilon\lesssim 100$. Nevertheless, since the ratio
$v_i/v_m$ grows with $L^{4/3}$ and with $\sigma^{-1}$, $\epsilon$
may be much larger in other extensions of the SM. We have seen
that if this is the case, the electroweak baryon inhomogeneities
will probably affect the dynamics of the quark-hadron phase
transition, as discussed in Ref \cite{s03}.

For the size of the inhomogeneities, we found in the case of the
MSSM a value of order $10^{-6}-10^{-5}$ times  the age of the
Universe (in agreement with Ref. \cite{h95}). However, this
quantity is sensitive to $\eta$ and $\sigma$, and can deviate
significantly from this value in other models.

These behaviors are illustrated in Figs. \ref{figsize} and
\ref{figvivm}. Our analytical expressions agree with the numerical
computations within the order of magnitude. We have considered
parameters that varied in a wide range in order to cover possible
extensions of the SM, other than the MSSM. Furthermore, our
results can be easily extrapolated beyond these ranges.

\section*{Acknowledgements}

A. M. is grateful to M. Quir\'os for useful discussions, and to
the Physics and Mathematics Institute at Michoacana University,
where this work was initiated, for kind hospitality. The work by
F. A. was supported by CIC-UMSNH and Conacyt grant 32399-E.

\appendix

\section{The chiral asymmetry}

In this appendix we find an analytical approximation for the
density $n_L(z)$. In section 2 we have seen that this is of the
form \begin{equation} n_{L}=Ae^{-\omega _{s+}z} , \label{apnl}
\end{equation}
and we argued that $A\propto v_w$, which is all we need to know
for the present paper. However, this may not be true in general,
so it is important to see under which conditions the coefficient
$A$ depends only linearly on $v_w$.

As we have seen, the solution to Eq. (\ref{difus}) is given by
Eqs. (\ref{nlgral}) and (\ref{omega}), with different values of
$\Gamma$, and consequently of $\omega_\pm$, $A$, and $B$ in the
broken and symmetric phases. Specifically,
$\Gamma_b=\tilde{\Gamma}$, $\Gamma_s=a\Gamma_{ws}$, where the
index $b$ ($s$) stands for the broken (symmetric) phase. According
to Eq. (\ref{omega}) we then have $\omega _{i\pm }=v_{w}/2D\pm
\left[\left( v_{w}/2D\right) ^{2}+\Gamma_i /D\right]^{1/2}$, and
the constants $A_i$ and $B_i$ are given by the boundary conditions
as explained in section \ref{baryo}. As a result, one obtains
\begin{eqnarray}
A_{b} &=&\left[ D\left( \omega _{b+}-\omega _{b-}\right) \right]
^{-1}\int_{-\infty }^{0}e^{\omega _{b+}z}\gamma\left( z\right)
dz,  \notag
\\
B_{s} &=&\left[ D\left( \omega _{s+}-\omega _{s-}\right) \right]
^{-1}\int_{0}^{\infty }e^{\omega _{s-}z}\gamma \left( z\right) dz,
\label{coef1}
\end{eqnarray}
and
\begin{eqnarray}
A_{s} &=&\frac{\omega _{b+}-\omega _{b-}}{\omega _{s+}-
\omega_{b-}}A_{b}-\frac{\omega _{s-}-\omega
_{b-}}{\omega_{s+}-\omega_{b-}}B_{s},
\notag \\
B_{b} &=&\frac{\omega _{b+}-\omega _{s+}}{\omega _{s+}-\omega
_{b-}}A_{b}+ \frac{\omega _{s+}-\omega _{s-}}{\omega _{s+}-\omega
_{b-}}B_{s}. \label{coef2}
\end{eqnarray}

The source $\gamma $ is proportional to the wall velocity, $\gamma
=v_{w}\tilde{\gamma} $. If $\tilde{\gamma}\left( z\right) $ is
localized in some region around the wall, then it can be easily
seen that the coefficients (\ref{coef1}) are such that outside
this region only the exponential with the right sign of $ \omega $
survives in each phase. To obtain an analytical result, we need to
use an approximation for $ \tilde{\gamma}$. The simplest
approximation for a function $\tilde{\gamma}\left( z\right) $ that
is localized inside the wall is a step function \cite{hn96}, say,
$\tilde{\gamma}=\tilde{\gamma}_0$ for $-L_{w}<z<0$, and
$\tilde{\gamma}=0$ outside. (The value of the constant
$\tilde{\gamma}_0$ depends on the $CP$ violating force inside the
wall.) Then, the solution for $z>0$ is of the form (\ref{apnl}),
with
\begin{equation}
A=\frac{\left( 1-e^{-\omega _{b+}L_{w}}\right)
v_{w}\tilde{\gamma}_0}{D\omega _{b+}\left( \omega _{s+}-\omega
_{b-}\right) }. \label{eq1}
\end{equation}

Notice that for $\tilde{\Gamma} \sim 10^{-1}T$, the relation
$\left( v_{w}/2D\right) ^{2}\ll \Gamma _{b}/D$ holds for any
possible value of the wall velocity, as long as $D\gtrsim
10T^{-1}$. So, the $\omega_{b\pm}$ are insensitive to the wall
velocity, $\omega_{b\pm}\simeq\pm \sqrt{\tilde{\Gamma}/D} $.
Furthermore, since $\Gamma_s\ll\Gamma_b$, we can also neglect
$\omega_{s+}$ in Eq. (\ref{eq1}), and the coefficient $A$ becomes
\begin{equation}
A\simeq v_{w}\frac{\tilde{\gamma}_0\left(
1-e^{-L_{w}\sqrt{\tilde{\Gamma} /D}}\right)}{\tilde{\Gamma}} ,
\end{equation}
which is evidently linear in $v_{w}$. This expression for the
coefficient $A$ coincides with the one obtained in Ref.
\cite{hn96}. On the other hand, the $z$-dependance in Eq.
(\ref{apnl}) is different due to our inclusion of the sphaleron
process in the diffusion equation for $n_L$. As we have seen, this
modification does not introduce any qualitatively different
behavior in the final result for the baryon number density.
However, some quantitative $\mathcal{O}(1)$ difference in the BAU
shows up for $v_w^2 \lesssim D\Gamma_{ws}$, i.e., for $v_w
\lesssim 10^{-2}$.

\end{document}